\documentclass[12pt,epsfig,amsmath]{article}

\usepackage{graphicx}
\usepackage{dcolumn}
\usepackage{bm}
\usepackage{verbatim}
\usepackage{comment}
\usepackage{amsmath}

\usepackage{graphicx}
\usepackage{epsfig}
\usepackage{picinpar}
\usepackage{wrapfig}
\usepackage{comment}

\usepackage{hyperref}

\renewenvironment{thebibliography}[1]
        {\begin{list}{\arabic{enumi}.}
        {\usecounter{enumi}\setlength{\parsep}{0pt}
         \setlength{\itemsep}{0pt}
         \settowidth
        {\labelwidth}{#1.}\sloppy}}{\end{list}}

\author{Mathieu Ribordy\footnote{{\it Email Address:} mathieu.ribordy@umh.ac.be}\\
{\small Universit\'e de Mons-Hainaut, 19 Av. Maistriau, 7000 Mons, Belgium
}}
\title{Reconstruction of Composite Events \\ in Neutrino Telescopes}
\date{}

\begin{document}
\maketitle
\thispagestyle{empty}

\vskip2cm
\begin{abstract}
Neutrino telescopes detect the emission of Cherenkov light resulting from the tracks and showers
of charged and neutral current neutrino interactions. These tracks or showers are reconstructed using a corresponding probability density function (PDF) which depends on measured time and location of the detected photons.
We call a composite event the mixed detector response due to the juxtaposition of more than one Cherenkov light source (track or shower). This paper presents the construction of a generic PDF corresponding to a composite hypothesis.
This composite PDF is therefore useful to reconstruct an arbitrary event topology and to favor or discard a given event topology hypothesis.
\end{abstract}

\section{Introduction}
\setcounter{equation}{0}
Neutrino telescopes focus on the reconstruction of single neutrino-induced muon tracks or single showers from neutral and charged current neutrino interactions. There are however events with a more complex topology due to a superposition of Cherenkov photons emitted from more than one track and/or shower.
There are several reasons 
to develop the ability of reconstructing composite events (referred to as composite event reconstruction in the following),
for instance:
in a cubic km neutrino telescope, down-going uncorrelated muon (originating from distinct atmospheric showers) track events happen at a rate of a few per second at 2 km depth in seawater or ice; 
exotic channels may result in nearly parallel track events (production from neutrino-nucleon interactions of $\tilde{\tau}$ pairs in supersymmetric extensions of the standard model~\cite{staus} or of muon pairs from micro black hole evaporation in models with extra-dimensions~\cite{mubh});
PeV $\tau$ double bang events with significant consequences for particle physics (one expects 0.1-10 resolved events per year in IceCube~\cite{conv_tig}).
Furthermore, the prolific number of shower-track events from muons undergoing catastrophic energy losses or from neutrinos converting in the detector effective volume concur with the composite hypothesis.
The dedicated method of composite event reconstruction presented here may not be well adapted to the specific topology of muon bundle events: in which case, the fading light output along the bundle axis due to decreasing muon density is the important ingredient in the determination of the bundle properties (primary cosmic ray composition studies~\cite{ama_spase}).

In section~\ref{sct:singlesource}, single source reconstruction and likelihood definitions are briefly reviewed. 

Section~\ref{sct:composite} considers the case of a composite event topology and demonstrates the combination of source PDF's into a generic multisource PDF. In contrast to the single source PDF, the composite PDF explicitly depends on energy; the contributions from the sources to the detected light pattern is contingent upon the relative weights of each source.
Thus, the reconstruction of the directional parameters of a composite event implies the reconstruction of the energy parameters.
The composite PDF extends our current ability to reconstruct single topological events.

The reconstruction of composite events is performed in Section~\ref{sct:usecase} with the presentation of an extreme case, reconstructing simultaneously propagating parallel tracks in the IceCube array~\cite{icecube}. The reconstruction of an event is the minimization process of attaining the best likelihood given a topology hypothesis. We intend to demonstrate that the comparison of likelihoods which correspond to double muon and single muon event hypothesis may be used to favor one topology over another.

The normalized PDF describing the arrival time of a single photon originating fron source $k$ at receiver $j$, given it has been detected, is written  $p^{(k,j)} (t_i) \equiv p_i^{(k,j)}$.
In the case of the detection of $N_j$ photons at times $\{t_1,...,t_{N_j}\}$, the corresponding PDF is written $p^{(k,j)} (t_1,\,...,\,t_{N_j})$. It can be derived from the single photon PDF through the relation $p^{(k,j)} (t_1,\,...,\,t_{N_j}) = N_j!\,\prod_i^{N_j} p^{(k,j)}_i$. The factor $N_j!$ reflects the indistinguishability of photons (any of the $N_j!$ photon permutations leads to the same PDF and should therefore be summed up).

The probability to detect $n$ photons from source $k$ at receiver $j$ is written $f^{(k,j)}_n$ and will be assumed to obey Poisson law $f^{(k,j)}_n = e^{-\mu}\mu^n/n!$, where $\mu \equiv\mu^{(k,j)}$ is the mean number of detected photons (oor hits or photo-electrons).

The $p^{(k,j)}_i$'s and $\mu$'s depend implicitly on the relative orientation and distance between the source and the receiver, the optical properties of the medium, the receiver efficiency, etc.
The $\mu$'s additionally depend on the source intensity (related to the source energy).

Consider now $K$ sources and the detection of $N_j$ photons at times $\{t_1,...,t_{N_j}\}$ at receiver $j$. The corresponding normalized composite PDF 
$$\tilde{p}^{\{K,j\}}_{N_j} \equiv \tilde{p}^{\{K,j\}}(t_1,\,...,\,t_{N_j})$$ 
will be constructed from the $f^{(k,j)}_n$'s and the $p^{(k,j)}_i$'s defined above.
We also conveniently define the composite density $p_{N_j}^{\{K,j\}} = N_j!\,\tilde{p}^{\{k,j\}}_{N_j}$ used in the reconstruction formulas.

\section{Single source reconstruction}\label{sct:singlesource}
For illustration purposes, we consider a source which can be parametrized by giving a vertex $\vec{q}$, the incidence angles $\theta$, $\phi$ and its energy $E$ (e.g. infinite tracks or showers), that is $n_{\mathrm{d.o.f}}=6$. For a source with a different topology, e.g. an extended source or a finite track, the parameters should be defined accordingly and $n_{\mathrm{d.o.f}}$ acquire the corresponding value; this will not affect the results below.
The reconstruction of the parameters consists in maximizing the log likelihood expression for an arbitrary hit time series for each receiver~\cite{amanda_reco}:
\begin{equation}
\label{llhE}
\ln{\cal L} = \frac{1}{n_{\mathrm{hit}}-n_{\mathrm{d.o.f}}}  \Bigg(\sum_{j=1}^M \Big(
\ln{p^{\{1,j\}}_{N_j}} 
+ \ln{f^{\{1,j\}}_{N_j}} \Big)
+ \sum_{\{j|N_j=0\}} \ln{f^{\{1,j\}}_0}\Bigg)
\end{equation}
where $j \in \{1,\,..,\,M\}$ is the hit receiver index and $n_{\mathrm{hit}}=\sum_j N_j$.
The last term represents the information from receivers which have not detected any photons. This term can be dropped, when the calculation speed is a concern, but will consequently be accompanied by a degradation in the precision of the reconstructed parameters.
We notice that eq.~(\ref{llhE}) can be split in two parts depending separately on energy $E$ and directional ($\{\vec{q},\,\theta,\,\phi\}$) parameters. This leads to introduce the reduced log likelihood formulas:
\begin{equation}
\label{llh}
\ln{\cal L}_{\mathrm{dir}} = \frac{1}{n_{\mathrm{hit}}-n_{\mathrm{d.o.f}}} \sum_{j=1}^M 
\ln{p^{\{1,j\}}_{N_j}} 
\end{equation}
and
\begin{equation}
\label{llh_onlyE}
\ln{\cal L}_E = \frac{1}{n_{\mathrm{hit}}-n_{\mathrm{d.o.f}}}  \Bigg(\sum_{j=1}^M
\ln{f^{\{1,j\}}_{N_j}} + \sum_{\{j|N_j=0\}} \ln{f^{\{1,j\}}_0}\Bigg)
\end{equation}
where $n_{\mathrm{d.o.f}}$ take the corresponding values (respectively 5 and 1). Expressions~(\ref{llh}) and~(\ref{llh_onlyE}) can be used to reconstruct the corresponding subset of parameters in a first approximation, when calculation speed is a concern. However, best results are obtained by using expression~(\ref{llhE}), which more fully exploit the available information.

\section{Generic composite event topology reconstruction}\label{sct:composite}
In the case of a composite event, recorded hits do not provide the information from which source they originate. So, in this section, we consider a case with an arbitrary number of sources and an arbitrary number of hits to each receiver, thereby describing the construction of the generic PDF $p_{N_j}^{\{K,j\}}$ given the sets ($k\in\{1,\,...,\,K\}$):
$$\{ p^{(k,j)}_i\}_{1\le i\le N_j}\,\mathrm{and}\,\{ f^{(k,j)}_n \}_{0\le n\le N_j}.$$ 
We also exhibit the log likelihood expression to use for reconstructing composite events.

\subsection{Two-source composite}
The simplest composite event consists of two sources and the detection of one photon at receiver $j$, the PDF is given by
\begin{equation}
\tilde{p}^{\{2,j\}}_1 = K^{{\{2,j\}}}_1 \Big[
f^{(1,j)}_1 \,p^{(1,j)}_1\, f^{(2,j)}_0 + f^{(1,j)}_0\, f^{(2,j)}_1 \,p^{(2,j)}_1
\Big].
\end{equation}
where we apply a superposition principle for the photon, which originates from source $1$ but not from source $2$ with corresponding weights $f_1^{(1,j)}$ and $f_0^{(2,j)}$, and vice-versa.
The normalization is given by integrating over all hit times\footnote{This is worth mentioning, as it has not been accounted for in the past, that experimentally, the recording time window for one event has a finite width. This fact should be adequately reflected in the hit probabilities $f^{(k,j)}_n$ and the basic PDF $p^{(k,j)}(t)$, which should be renormalized within this time window.}, meaning that the hit was actually measured:
$$
\int dt_{1} \tilde{p}^{\{2,j\}}_{1} =1 \longrightarrow K^{\{2,j\}}_1 = \Big(f^{(1,j)}_1\,f^{(2,j)}_0 + f^{(1,j)}_0\,f^{(2,j)}_1\Big)^{-1}.
$$

For two photons, we have
\begin{equation}\label{eq:twophotons}\begin{split}
\tilde{p}^{\{2,j\}}_2 = K^{{\{2,j\}}}_2 \Big[&
2\,f^{(1,j)}_2\,p^{(1,j)}_1\,p^{(1,j)}_2\,f^{(2,j)}_0
+ f^{(1,j)}_1\,f^{(2,j)}_1\,\big(p^{(1,j)}_1\,p^{(2,j)}_2 + p^{(1,j)}_2\,p^{(2,j)}_1\big) \\
&+ 2\,f^{(1,j)}_0\,f^{(2,j)}_2\,p^{(2,j)}_1\,p^{(2,j)}_2 
\Big]
\end{split}\end{equation}
where combinatoric factors appear to account for the indistinguishability of the photons.


This is generalized for an arbitrary number of hits:
\begin{equation}\label{eq:P2}\begin{split}
\tilde{p}^{\{2,j\}}_{N_j} = K^{{\{2,j\}}}_{N_j} \times
\sum_{m=0}^{N_j} \Big(& (N_j-m)!\,f^{(1,j)}_{N_j-m} \,m!\, f^{(2,j)}_m 
\times \sum_{\{\pi_j\}} p^{(1,j)}_{\pi_1}... p^{(1,j)}_{\pi_{N_j-m}}p^{(2,j)}_{\pi_{N_j-m+1}}...p^{(2,j)}_{\pi_N}\Big)
\end{split}\end{equation}
where the sum over $\{\pi_j\}$ is the sum over $C(N,m)=N!/((N-m)!\,m!)$ different combination terms.
$K^{{\{2,j\}}}_{N_j}$ is extracted from the normalization condition
$$
\int dt_1\,...\,dt_{N_j}\, \tilde{p}^{\{2,j\}}_{N_j} = 1  
\longrightarrow 
K^{{\{2,j\}}}_{N_j} = \Big({N_j}!\,\sum_{m=0}^{N_j} \,f^1_{{N_j}-m} \, f^2_m \Big)^{-1}
$$

\subsection{$K$-source composite}
Expression~(\ref{eq:P2}) is generalized for an arbitrary number of $K$ sources in the composite:
\begin{equation}\label{eq:PN}
\tilde{p}^{\{K,j\}}_{N_j} = K^{\{K,j\}}_{N_j}
\sum_{\substack{\{k_1,\,...,\,k_{N_j}\}\\ 1\le k_j\le K}}
\Bigg(
\Big[\prod_{i=1}^{N_j} p_i^{(k_i,j)}\Big]
\Big[\prod_{k=1}^K\,
f^{(k,j)}_{\sum_{i=1}^{N_j} \delta_{k_i}^k}
\big(\sum_{i=1}^{N_j} \delta_{k_i}^k\big)!
\Big]
\Bigg)
\end{equation}
$\tilde{p}^{\{K,j\}}_{N_j}$ is written as a sum of $K^{N_j}$ \{source - hit\} configurations instead of a permutation over hit indices as in eq.~(\ref{eq:P2}). 
Eq.~(\ref{eq:PN}) should be used also when $K=2$ (instead of eq.~(\ref{eq:P2})) as it reduces the number of terms (thus reducing the computation time).
The normalization is given by
\begin{equation}
K^{\{K,j\}}_{N_j} =\Bigg(
\sum_{\substack{\{k_1,\,...,\,k_{N_j}\}\\ 1\le k_j\le K}}
\Big[\prod_{k=1}^K\,
f^{(k,j)}_{\sum_{i=1}^{N_j} \delta_{k_i}^k}
\big(\sum_{i=1}^{N_j} \delta_{k_i}^k\big)!
\Big]
\Bigg)^{-1}
\end{equation}

This formulation for a composite PDF is independent of the specific expressions for $p_i^{(k,j)}$ and $f_n^{(k,j)}$ (different $k$ may correspond to different source topologies).

\subsection{Likelihood formulation}\label{sct:genllh}
The calculation of the generalized log likelihood expression~(\ref{llhE}) requires 1) using the composite PDF defined above and 2) setting appropriately for the increased number of free parameters $n_{\mathrm{d.o.f}}$.
That is, substituting the one source expressions for the hit probabilities and the photon arrival time PDF with the corresponding composite expressions: 
$$f^{\{1,j\}}_0 \rightarrow f^{\{K,j\}}_0\,\,\,\,\mathrm{and}\,\,\,\,P_{N_j}^{\{1,j\}}\rightarrow P_{N_j}^{\{K,j\}},$$ 
where 
$f^{\{K,j\}}_0 = \prod_{k} f^{(k,j)}_0$ and
$$P_{N_j}^{\{K,j\}} = (K^{\{K,j\}}_{N_j})^{-1} p_{N_j}^{\{K,j\}}$$ 
are the $M$ unnormalized PDF's which absorb the hit probabilities in their definition. 
The sets 
$$\{f_n^{(k,j)}\}_{k\in\{1..K\}}\,\mathrm{and}\,\{p_i^{(k,j)}\}_{k\in\{1..K\}}$$ are determined in accordance with the underlying topology of the $K$ sources defining the composite PDF. This reads:
\begin{equation}
\label{fullLLH}
\ln{\cal L}^{\{K\}} = \frac{1}{n_{\mathrm{hit}}-n_{\mathrm{d.o.f}}} \Big(
\sum_{j=1}^M 
\ln{P^{\{K,j\}}_{N_j}} + 
\sum_{\{j|N_j=0\}} \ln{f^{\{K,j\}}_0}
\Big) 
\end{equation}

The explicit dependence on energy with the inclusion of the $f_n^{(k,j)}$ terms in the definition of $P_{N_j}^{\{K,j\}}$ forbids the splitting of expression~(\ref{fullLLH}) in two distinct components: the reconstruction of the directional parameters cannot be disentangled from the energy parameter.
This is an important difference between the single and the composite source likelihood formulations: in the former, while it is admissible to reconstruct the energy parameter separately from the directional parameters, c.f. eq.~(\ref{llhE}) (because the energy-dependent $f^{(k,j)}_n$ in eq.~(\ref{eq:PN}) are absorbed in the normalization for $K=1$), in the latter ($K>1$), these factors are intertwined (the $f^{(k,j)}_n$ explicitly enter the definition of the $P^{\{K,j\}}_N$ preventing the splitting of $\ln{\cal L}^{\{K\}}$).

These notations have the advantage of unifying the log likelihood formulation: the eq.~(\ref{fullLLH}) is expressed exactly like eq.~(\ref{llhE}), the information regarding the actual number of sources in the composite is contained in $f^{\{K,j\}}_0$ and $P_{N_j}^{\{K,j\}}$. The single source log likelihood formulation is therefore recovered by setting $K=1$.

For instance, in order to reconstruct two muon tracks, the expression (\ref{fullLLH}) which depend on 12 free parameters should be maximized. In the case of two parallel muon tracks traveling simultaneously, this number is reduced to 9 ($\{\vec{q_1},\,\theta,\,\phi,\,E_1,\,E_2,\,d_{\mathrm{sep}},\,\eta\}$ where $d_{\mathrm{sep}}$ and $\eta$ are respectively the distance between the two muon tracks and the angle of rotation of one muon relative to the other).

\subsection{Inclusion of noise in the composite PDF}
When the considered topology hypothesis consist in more than one source,
the individual noise of the receivers and the environmental noise levels must be included as a new source and cannot be included in the single source PDF. Thereby the number of sources is increased by one.

\section{Use Case}\label{sct:usecase}
We discuss in this section a use case prompted by a study of the evaporation of a micro black hole into a muon pair~\cite{mymubh} (see~\cite{mubh} for related micro black hole phenomenology studies). This use case also applies to other exotic channels (e.g. $\tilde{\tau}$ pairs~\cite{staus}).

\subsection{Motivations}
A fraction of the micro black holes produced in the neighborhood of the detector evaporate into PeV muon pairs, among other particles, which will travel 10's of km before reaching the detector with typical energies around 3 TeV.
These events are characterized by nearly parallel and simultaneously traveling muons. Their incidence direction distribution is peaking horizontally. 
The rate of resolved events traversing the IceCube~\cite{icecube} detector will depend on the low energy gravity scale parameters (such as the number of extra-dimensions and the Planck scale) as well as on the detector resolution power (driven by the medium properties and the spacing of the detecting devices). The latter is the object of this study.
The background rate of horizontal atmospheric neutrino-induced muons is approximately 100 km$^{-2}$ yr$^{-1}$ above 1 TeV, within one degree from the horizon (see e.g. Lipari~\cite{conv_lip} or Tig~\cite{conv_tig} for conventional atmospheric neutrino flux parametrizations).
This use case is a challenge for the application of the composite reconstruction, because of the conjunction of simultaneity, parallelism and closeness of the muons.

\subsection{Event generation}
The AMANDA simulation software AMASIM~\cite{amasim}, including the depth-dependent ice properties, adapted for the IceCube geometry has been used to generate 1000 single muon track events and 1000 double muon track events. 
Subsequently the directional parameters were reconstructed by means of a first guess method (direct walk)~\cite{amanda_reco}, providing the seed for the log likelihood reconstructions, which were performed under the single muon track and the double muon track topology hypotheses.

Parallel muon pairs were generated 3 km away from the detector center with $E_\mu=10$ TeV such that they mostly cross completely the detector active volume: -100 m $<z<$ 100 m, $-\pi<\phi<\pi$, $-\pi<\eta<\pi$, 5 m $<d_{\mathrm{sep}}<$ 250 m, -200 m $<xy<$ 200 m, where $z$ is a vertical shift for the first muon from the detector center, $xy$ is a horizontal shift for the detector center in the $xy$-plane, $\phi$ is the incidence angle in the $xy$-plane; $d_{\mathrm{sep}}$ and $\eta$ were defined section (\ref{sct:genllh}).
99.5\% of the generated muon pairs did pass a majority trigger of 24 fired receivers (photomultipliers) within 2.5 $\mu$s), each muon having a mean energy of 3 TeV when reaching the point of closest approach from the detector center.

The energy at the detector and incidence direction distributions of the generated events reproduce only the main features of the aforementioned detailed study. This simplification of the physical case nonetheless provides a useful framework with which to test the capabilities of the composite reconstruction.

\subsection{Normalized PDF}
An explicit expression for the PDF~\cite{pandel} (also discussed in~\cite{amanda_reco}) which can be used in shower and track reconstruction is
\begin{equation}
\label{pandel}
p(\xi,\,\rho,\,t)\,=\, \frac{1}{\Gamma(\xi)}\,\rho^{\xi}\,t^{\xi-1}\,e^{-\,\rho t} 
\end{equation}
Given a detected photon at a distance source-receiver $d$, this expression allows to calculate the probability of a given time delay $t$ (actual arrival time minus geometrical time) of the photon. 
$\xi\equiv d/\lambda,\,\rho\equiv 1/\tau +c/\lambda_{\mathrm{abs}}$, where the variables $\lambda,\,\tau,$ and $\lambda_{\mathrm{abs}}$ characterize the medium. This is described in details~\cite{us1}. Variant to this expression, accounting for a gaussian time response of the receivers~\cite{us1} and for variable optical ice properties w.r.t. to the depth~\cite{us2} were used for the reconstruction (it can be shown that using the average of the absorption and scattering coefficients between the photon emission location and the receiver is a valid approximation).

\subsection{Hit probability}
To calculate the relative probability of a photon from one track w.r.t. another, the hit probability must be calculated. First, we express the unnormalized version of eq.~(\ref{pandel}), that allows to calculate the probability for a photon to reach and to be detected at some distance $d$ with some time delay $t$ in an absorbing medium:
\begin{equation}
\label{P}
P(\xi,\,t)\,=\, \frac{1}{\Gamma(\xi)}\,\frac{1}{t}\Big(\frac{t}{\tau}\Big)^{\xi}\,e^{-\frac{t}{\tau}-\frac{(d+c_{\mathrm{ice}}t)}{\lambda_{\mathrm{abs}}}} 
\end{equation}
so that we can derive the light intensity at distance $d$ by integrating~(\ref{P}):
\label{W}
\begin{equation}
W(d)=\int_{t=0}^{\infty} P(\xi,\,t) dt = \Big[e^{1/\lambda_{\mathrm{abs}}} (1+c_{ice}\tau/\lambda_{abs})^{1/\lambda}\Big]^{-d} \equiv w^{-d}.
\end{equation}

The explicit depth dependence of $p$, $P$, $W$ and subsequent derived quantities is omitted below. It must be noted that this depth dependence is not negligible for a medium with varying optical properties in general and for the medium where IceCube 
modules are deployed
(South Pole ice~\cite{ice_paper}) in particular.

The average number $\mu(d,\,\theta,\,E)$ of detected photons is proportional to $W(d)$ (for an ideal receiver), with an absolute normalization factor $a(\theta,\,E)$. The factor $a$ accounts for the photomultiplier quantum efficiency and for the relative orientation to the track $\theta$.
The average number of photons is expressed as
$$
\mu(d,\,\theta,\,E) = a(\theta,\,E)\, w^d + b_{\mathrm{noise}}
$$
where $b_{\mathrm{noise}}$ is the detector/environment noise contribution (note that the split is approximative and $a$ may also depend on $d$).
In the following we drop the $\theta$-dependency 
(however this may not be a satisfying approximation for short distances between the source and the receiver, as Cherenkov photons have not been isotropized through scattering).

\begin{figure}[h]
\centering
\epsfig{file=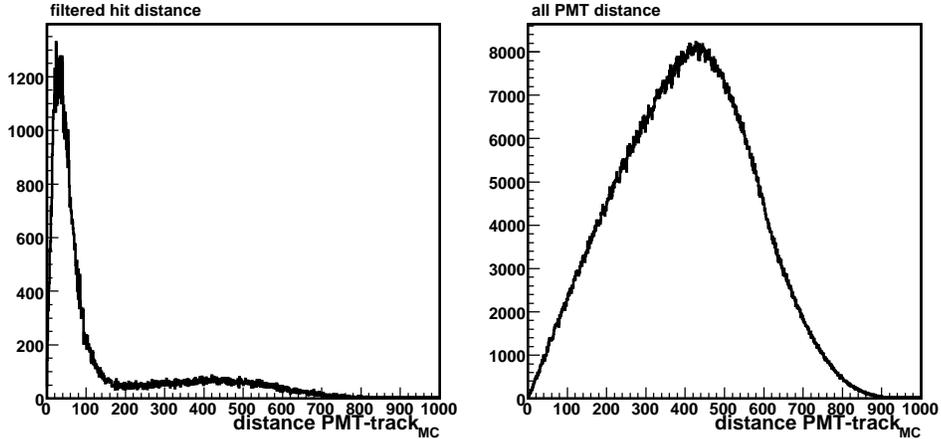,width=.8\textwidth}
\caption{Filtered hit and receiver perpendicular distances to MC track.}
\label{fig:phitA}
\end{figure}

\begin{figure}[h]
\centering
\epsfig{file=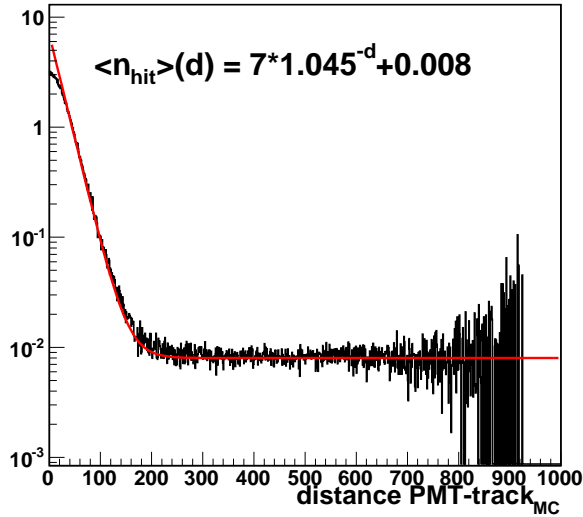,width=.5\textwidth}
\caption{Resulting average number of hits per receiver w.r.t. the distance from Fig.~(\ref{fig:phitA}). The hit probability can be extracted: $p_\mu(n\geq1)=1-e^{-\mu(d)}$, where $\mu(d)=a\,w^{-d}+b=a\,W(d)+b$. The fitted parameters correspond to $>$50\% and $\approx$10\% probability of having at least one signal hit at resp. 50 m and 100 m distances.}\label{fig:phitB}
\end{figure}

Fig.~\ref{fig:phitA} shows the distributions of the cumulated number of hits (left) and receivers (right) for all the single muon events in the sample w.r.t. the perpendicular distance to the generated single muon track.
Dividing the first distribution by the second results into the distribution of the average number of hits, shown Fig.~\ref{fig:phitB}, from which $\mu$ is now extracted\footnote{Alternatively, $\mu(d)$ can be extracted using the distribution of hit receivers instead of the distribution of the cumulated number of hits.} by fitting $\mu(d) \,=\, a\,w^{-d} + b$. We obtain the proportionality factor $a\approx7$, the noise contribution $b\approx0.008$ and $w\approx1.045$.
When we calculate $w$ from the usual medium PDF parameters~\cite{amanda_reco}, we get a close agreement
($w=1.044$ using $\tau=557\,\mathrm{ns},\,\lambda_{\mathrm{eff}}=25\,\mathrm{m},\,\lambda_{\mathrm{abs}}=98\,\mathrm{m}$). This serves as an independent confirmation that the result above is correct.
In general, $a$ is a function of energy (or even $a=a(\theta(d),\,E)$). In this specific study, $a$ is held constant, as the muons all belong to the same energy range. 

To reconstruct the energy (or more exactly to make a calorimetric measurement of the energy deposit in the detector), it would suffice to find a parametrization (using the Monte Carlo) for $\mu(d)$ as a function of $E$ as well\footnote{The energy resolution $\log{(E_{\mathrm{rec}}/E_{\mathrm{MC}})}$ for horizontal muons (using the MC described above) crossing the detector with energies between 1 TeV and 1 PeV is approximately 20\% with this method. However, with the actual digital IceCube array, the energy can be reliably reconstructed up to much higher values.}.

The hit probabilities $f_n^{(2,j)}$ can now be calculated from $\mu(d)$, so that the normalized $p_i^{\{2,j\}}$ can also be computed for the full likelihood reconstruction.

\subsection{Reconstruction and favored hypothesis}
Reconstruction has been performed using eq.~(\ref{fullLLH}) for single and double track hypotheses with fixed energy ({\it i.e.} $\mu=\mu(d)$). 
A dedicated receiver response filter did consist in rejecting photomultipliers with more than four hits within 8 $\mu$s around the trigger time or with hit response inconsistent with a single photo-electron (within the same time window).
The quality of this filter can be understood by looking at Fig.~\ref{fig:phitB}. At large distances (more than few 10's of m), the probability of having more than 4 hits is negligible. At short distances (less than few 10's of m), this probability is no longer negligible, but the fit no longer reproduces the data. This is mainly due to the fact that at short distances, hits can be the results of overlapping, and thus unresolved photo-electrons (the MC simulates the old AMANDA data acquisition system). This introduces a bias affecting the single and muon pair event reconstruction. However the bias is expected to be mild, given the muon energies.

\begin{figure}[h]
\centering
\epsfig{file=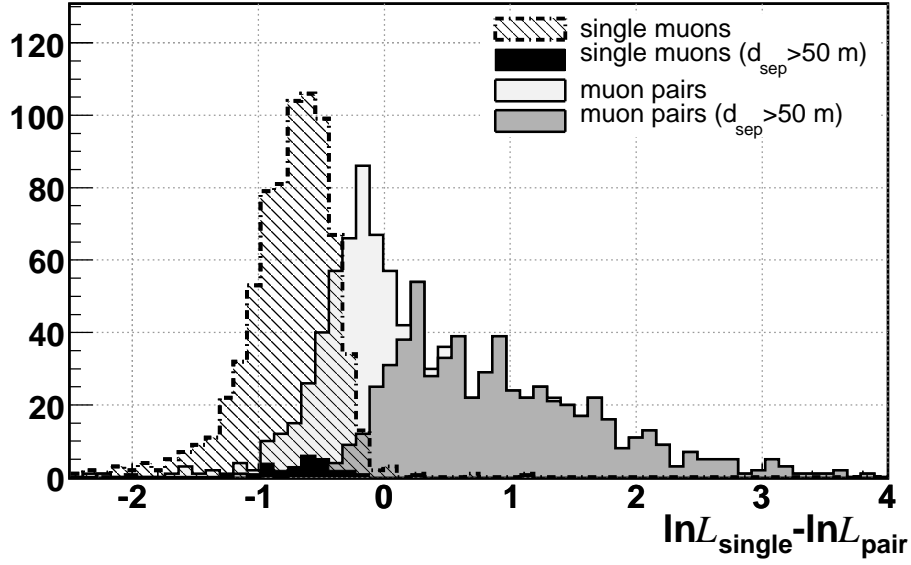,width=.8\textwidth}
\caption{\label{fig:ssd_llhA}
Distribution of the difference between single and double muon log likelihood reconstruction hypotheses $\ln{\cal L}_{\mathrm{single}}-\ln{\cal L}_{\mathrm{pair}}$ for single muon and muon pair events (with and without a selection on the reconstructed $d_{\mathrm{sep}}$).
}
\end{figure}

\begin{figure}[h]
\centering
\epsfig{file=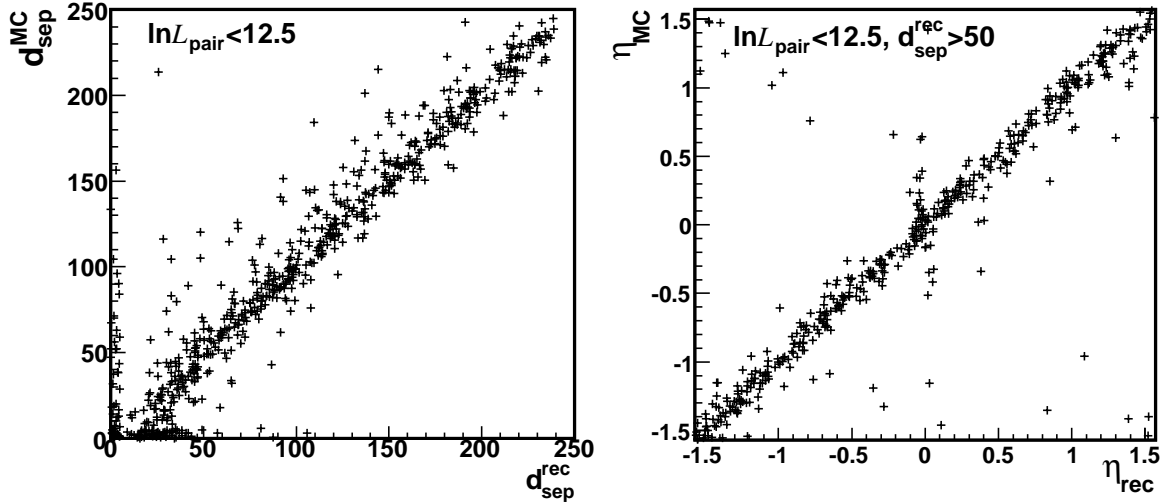,width=1\textwidth}
\caption{\label{fig:ssd_llhB}
Left: Correlation plots of generated VS reconstructed $d_{\mathrm{sep}}$. Only events with $\ln{\cal L}_{\mathrm{pair}}<12.5$ are selected.
Right: Correlation plots of generated VS reconstructed $\eta$ (for $d>50$ m). Only events with $\ln{\cal L}_{\mathrm{pair}}<12.5$ are selected.}
\end{figure}

Some of the results are shown Fig.~\ref{fig:ssd_llhA} and~\ref{fig:ssd_llhB}:
\begin{itemize}
\item Fig.~\ref{fig:ssd_llhA}: the distribution of the difference between single and double muon log likelihood reconstruction hypotheses for single muon and muon pair events clearly demonstrates how an hypothesis can be favored against another. Notice the remarkable discriminating power between single muon and muon pair events, when $d_{\mathrm{sep}}^{\mathrm{rec}}>50$ m.

\item Fig.~\ref{fig:ssd_llhB}: the plots illustrate the correlation between reconstructed and generated $d_{\mathrm{sep}}$ (left) and $\eta$ (right). Though no first guess reconstruction was used to set initial conditions for $d_{\mathrm{sep}}$ and $\eta$, the correlation is nevertheless very strong (the correlation factors are respectively 97\% and 87\% for $d_{\mathrm{sep}}$ and $\eta$ ($d_{\mathrm{sep}}>50$ m)).
\end{itemize}

Evidently, seawater neutrino telescopes would improve the resolution to shorter $d_{\mathrm{sep}}$ given the longer scattering length of the medium.

\section{Conclusion}
We showed how to construct a multisource PDF by combining single source PDF's and how to formulate the corresponding likelihood, thus enabling composite event reconstruction.

Then using a IceCube simulation, we demonstrated that events were successfully reconstructed in the difficult case of close parallel muons.

While the construction of this PDF is completely generic, it may not be necessarily effective at high energy (e.g. PeV double bang events) or very close distances, when the number of photo-electrons becomes large which in turn increases the number of combinatoric terms in the composite PDF
(unless a compact approximation of~(\ref{eq:PN}) for large number of photo-electrons can be found).
However, it may reveal useful for the reconstruction of muons from distinct air showers and open experimental perspectives for various exotic searches.

\section*{Acknowledgment}
I am much indebted to G.~Japaridze for valuable discussions. I much thank P.~Toale for providing me with an extension of the AMANDA Monte Carlo adapted for the IceCube geometry and also P.~Herquet, F.~Grard and C.~P.~de~los~Heros for useful discussions. This work was supported by the Swiss National Research Foundation (grant PA002-104970).

\end{document}